\documentclass[sigconf,manuscript]{acmart}

\usepackage{wasysym}
\AtBeginDocument{%
  \providecommand\BibTeX{{%
    \normalfont B\kern-0.5em{\scshape i\kern-0.25em b}\kern-0.8em\TeX}}}

\setcopyright{iw3c2w3g}
\copyrightyear{2021}
\acmDOI{}

\acmConference[KidRec'21]{KidRec '21: 5th International and Interdisciplinary Perspectives on Children \& Recommender and Information Retrieval Systems (KidRec) Search and Recommendation Technology through the Lens of a Teacher- Co-located with ACM IDC 2021}{June 26, 2021}{Online Event}

\begin{document}
\title{All Together Now: Teachers as Research Partners in the Design of Search Technology for the Classroom}

\author{Emiliana Murgia}
\email{emiliana.murgia@unimib.it}
\affiliation{%
  \institution{Universit\`a degli Studi di Milano-Bicocca Milano}
  \country{Italy}
  }

 \author{Monica Landoni}
\email{monica.landoni@usi.ch}
\affiliation{%
  \institution{Universit\`a della Svizzera Italiana}
   \country{Switzerland}
  }

  \author{Maria Soledad Pera}
\orcid{0000-0002-2008-9204}
\email{solepera@boisestate.edu}
\affiliation{%
  \institution{PIReT - Dept. of Computer Science-- Boise State University}
  \country{United States}
  }
  
   \author{Theo Huibers}
  \authornote{Co-founder of Wizenoze.}
\email{t.w.c.huibers@utwente.nl}
\affiliation{%
  \institution{University of Twente}
  \country{The Netherlands}
  }

\renewcommand{\shortauthors}{Murgia, et al.}

\begin{abstract}
In the classroom environment, search tools are the means for students to access Web resources. The perspectives of students, researchers, and industry practitioners lead the ongoing research debate in this area. In this article, we argue in favor of incorporating a new voice into this debate: teachers. We showcase the value of involving teachers in all aspects related to the design of search tools for the classroom--from the beginning till the end. Driven by our research experience designing, developing, and evaluating new tools to support children’s information discovery in the classroom, we share insights on the role of the experts-in-the-loop, i.e., teachers who provide the connection between search tools and students. And yes, in our case we, always involving a teacher as a research partner.
\end{abstract}



\maketitle

\section{The Forgotten Stakeholder: The Teacher}

The integration of technology into classrooms has not been a smooth process due to barriers such as varying teacher perception of technology, uneven access to technology, and a lack of shared vision on how technology can best support classroom curriculum \cite{kearney2018teachers,domingo2016exploring}. Let us take, for example, \textit{search tools for the classroom}. Researchers and practitioners have studied for more than two decades how to design educational search tools explicitly tailored towards (young) students \cite{azzopardi2009puppyir,bilal2017analysis,druin2009children,jochmann2010children,pera2019empirical,landoni2019sonny,gossen2016search,wizenoze}. Still, one crucial concern remains unaddressed: the users (i.e., children in primary and secondary schools) and the designers (i.e., researchers and industry practitioners) emerge as the main stakeholders, while teachers are overlooked. The teacher, who serves as the expert-in-the-loop \cite{murgia2019will}, is another major general stakeholder, one that at the same time could be much more than that, as teachers have expertise in the domain they are teaching, in how to teach it, and in the way the children are learning. 

In 2019, along with international researchers and industry practitioners, we met at the $3^{rd}$ International and Interdisciplinary KidRec Workshop co-located with ACM IDC \cite{huibers20193rd}. Together, we crafted four important conditions for a \textit{good  search tool for the classroom}: (1) It provides resources that are logically relevant, useful, and foster learning, (2) It is designed with a user-centered perspective while acknowledging that multiple stakeholder perspectives and needs exist, (3) Users are deeply engaged with the system, and (4) It is ethically sound and supports the rights of the child \cite{huibers2020does}. Once again, the teacher did not emerge as a central player. 

Reflecting on our prior joint research experiences and lessons learned \cite{landoni2020we,landoni21ecce,aliannejadi2021children,umapSearchCompanionLBR,aliannejadi2020say,murgia2019will,murgia2019seven,pera2019little}, we posit that the perspective of the teacher is as indispensable as that the users and designers. Teachers' interpretations of the technical and educational needs inherent in the classroom context must guide each of the stages of development related to search tools for the classroom. Informed by the research work we conducted over the past four years about information retrieval tools tailored to primary school classrooms and anchored on education, human-computer interaction, and information retrieval disciplines, we offer observations on how to involve teachers in the research process \cite{murgia2019will,murgia2019seven}. We strongly suggest involving teachers as \textit{partners} in research, not just facilitators or study subjects themselves. And along the way, we describe the mutual benefits this collaboration brings to research.

\section{The Advantages of Involving Teachers}
As previously stated, one of our team members is a primary school teacher--an education expert specialized in the creative use of technology for teaching. Hence, user studies we have conducted to date in a classroom context have been designed, planned, and tested by naturally taking a teacher-centric approach. This has translated into non-intrusive studies that blend with regular classroom instruction enabling us, researchers, to capture authentic interactions and barriers experienced by young users when engaging with search tools. Moreover, we have ongoing collaborations with several primary school teachers in different schools in Italy, Switzerland, The Netherlands, and the United States. They have played varying roles in our studies, from informants and facilitators \cite{landoni2019sonny,landoni2019sonnyIDC,pera2019little} to administrators of the proposed search tasks and observers of children's behavior while performing the tasks \cite{aliannejadi2021children,landoni21ecce}. Recall that teachers are not only knowledgeable in the content they deliver, but also in the pedagogical side of their teaching and how children expect to use technology in the classroom. These are some of the reasons why involving teachers as \textit{partners in research} directly benefits research outcomes. We discuss some of the advantages we see for our team, which can extrapolate to other teams focused on advancing knowledge related to educational technology.

\begin{enumerate}
\item \textbf{The priorities of the research are well determined}. Instead of being guided by new research trends or cutting-edge technology, we let teachers--as the real experts--drive our explorations and outline the next steps on how to design ``good'' search tools for children in the classroom. In this instance, ``good'' (and even ``suitable'' for the classroom) is assessed by teachers. They are the experts in supporting children in the development of new skills, competencies, and knowledge, and specifically, in facilitating the search as learning paradigm \cite{collins2017search}. 

\item \textbf{Research design takes practice into account}. Educators joining research teams ensure that theory and practice go hand in hand, thus promoting a ``dynamic and effective teaching and learning process, boosting the digital inclusion of all individuals involved'' \cite{oliveira2018use}. Thinking about the practical use of technology is something we were fortunate to discuss early on as we set up our research agenda and enforced on each of our projects since \cite{landoni2020we}. More importantly, this is in line with the findings reported in a recently-published book by Daisy Christodoulou \cite{christodoulou2020teachers}, who draws on her classroom experience and from working in the education community. The author outlines a positive vision for the future, one where technology is developed in collaboration with teachers' expertise and ultimately used to improve educational outcomes for all.

\item \textbf{Research outcomes are usable}. \citet{mlekus2020raise} recently investigated characteristics of the user experience and potential determinants of technology acceptance. They conclude that technology that meets user experience (UX) expectations, is more likely to be accepted and used. Indeed, teachers have first-hand knowledge of the needs, expectations, and preferences of young searchers in the classroom, and thus can offer insights that can result in more swift adoption of newly-developed technology for the classroom. For instance, in our case, we relied on teachers to set up search tasks of growing complexity suitable for the classroom so that children would feel like engaging with and being constructively challenged by them. In turn, this allowed us to explore authentic interactions with search tools when presenting students with search tasks of varying complexity, as opposed to assuming that search tools can be evaluated in-depth in artificial search contexts.

\item \textbf{Findings make sense}. It was common practice to launch research focused on understanding children's search behavior by examining findings reported in the literature referring to adult searchers and then set up studies to determine the degree to which these would also hold when engaging children. Alternatively, it was also common to set up studies based on synthetic search tasks to establish initial premises to foster the design and development of search technology for children. By instead turning to  teachers, it is possible to better grasp how children are not simply ``short adults'' \cite{bilal2010mediated}. It is imperative to understand the ``relationship between teachers’ pedagogical beliefs and knowledge with the integration of technology to improve technology use in education'' \cite{najdabbasi2014integration}. In our case, teachers' insights allow us to make sense of study results not always in line with our expectations and outline new research paths. 

\item \textbf{It is much more fun \smiley}. The fun factor is a much-needed ingredient for the completion of any long-term research agenda. Collaboration among researchers with expertise in diverse research areas is not an easy feat; expectations and timelines are not always aligned, and more importantly, vocabulary across disciplines can be challenging to map \cite{mailloux2017benefits}. Rather than asking for occasional support to run user studies in the classroom, working together is much more fruitful and so much more fun. Being part of multidisciplinary teams help us to widen our vision and account for the needs and benefits of our target users: children in primary and secondary schools. 
\end{enumerate}

\section{The Benefits for the Teachers}
We have discussed the importance of building interdisciplinary research teams involving teachers when it comes to research pertaining to technology for the classroom. Grounded by our own research experience, we focused our discussion on search tools for the classroom. We noted early on that while there are numerous search tools, none of them are really standard across classrooms around the world \cite{azpiazu2017online}. We hypothesized that this could be due to a disconnect between the theory and the practice: how tools are designed--and more importantly requirements motivating said design--versus how tools are actually used on a regular basis. And then, who better to turn to address this gap than teachers, who are the nexus between the students and the tools themselves. Our experience thus far has demonstrated that partnering with teachers at every stage of the process brings many benefits to researchers. On the teachers' side, it offers them the opportunity to contribute in shaping from the very beginning the design of tools that respond to  children's needs and that can seamlessly be incorporated in the classroom to support student learning. By being part of research teams, teachers can also widen their vision of and perspective on the use of technology in the school context. Moreover, as opposed to focusing on ''busy work'', i.e., compulsory tasks assigned by school directors or ICT (Information and communications technology) managers, teachers can use their precious time outside of the classroom to focus on research projects that either serve as a bridge to address classroom technology gaps or let them acquire new knowledge to close the digital divide common to the classroom setting.

\begin{acks}
We would like to thank all the teachers involved in our research work thus far.
\end{acks}

\bibliographystyle{ACM-Reference-Format}
\bibliography{KidRec-References}

\end{document}